# High-precision tomography of ion qubits based on registration of fluorescent photons

Yu. I. Bogdanov, I.A. Dmitriev, B.I. Bantysh, N.A. Bogdanova, V.F. Lukichev

Valiev Institute of Physics and Technology of Russian Academy of Sciences, Moscow, Russia

**ABSTRACT**

We develop a new method for high-precision tomography of ion qubit registers under conditions of limited distinguishability of its logical states. It is not always possible to achieve low error rates during the readout of the quantum states of ion qubits due to the finite lifetime of excited levels, photon scattering, dark noise, low numerical aperture, etc. However, the model of fuzzy quantum measurements makes it possible to ensure precise tomography of quantum states. To do this, we developed a fuzzy measurement model based on counting the number of fluorescent photons. A statistically adequate algorithm for the reconstruction of quantum states of ion qubit registers based on fuzzy measurement operators is proposed. The algorithm uses the complete information available in the experiment and makes it possible to account for systematic measurement errors associated with the limited distinguishability of the logical states of ion qubits. We show that the developed model, although computationally more complex, contains significantly more information about the state of the qubit and provides a higher accuracy of state reconstruction compared to the model based on the threshold algorithm.

## 1. INTRODUCTION

It is well known that some computational tasks can be performed exponentially faster on a quantum processor compared to any modern or promising classical supercomputer [1]. The fundamental problem is to create a high-performance processor capable of executing quantum algorithms in an exponentially large computing space. Among the most promising platforms are systems based on ions in traps [2, 3], systems based on atoms in optical tweezers [4, 5], superconducting processors [6, 7], and photonic chips [8].

Currently, algorithms based on NISQ (Noisy Intermediate Scale Quantum) devices are of particular interest [9]. The use of trapped ions as a platform for NISQ calculations has the advantage of being able to perform native entangling operations between any pair of ions [10]. However, at present, ion states readout errors remain rather high [11–15]. At the same time, it is known that proper correction of readout errors in NISQ devices can significantly improve the accuracy of algorithm execution [16].

To date, the main readout mechanism for an ion qubit is the fluorescence process, which makes it possible to distinguish ion levels by the intensity of its glow after excitation [17-19]. The measurements make use of a photoelectron multiplier tube (PMT), which registers a certain number of counts caused by photon absorption over a fixed period of time. The traditional approach of discriminating states according to the PMT data is based on the threshold algorithm, when events with number of counts greater than or equal to a certain threshold value are referred to as bright states $|0\rangle$ [20]. This approach is inevitably associated with the presence of unaccounted systematic errors of two types. Firstly, the Poisson nature of photocount statistics results in mistaking the bright level for a dark one due to the fact that few photons will be registered. Secondly, the dark level can be mistaken for the bright one due to the finite time of amplitude relaxation, as well as the presence of scattered photons of the primary beam and the dark noise. There are various ways to correct such errors based on both hardware improvements and analysis of the detection time of photons [21-24].

It is difficult to completely get rid of systematic readout errors from a practical point of view. However, if the experiment requires the accumulation of statistical data for the same quantum circuit, then the use of a measurement model in the statistical reconstruction of the system parameters makes it possible to suppress systematic errors. This result can be achieved by using fuzzy measurement operators instead of standard projectors [25,26]. This, in turn, requires the preliminary construction of the measurement model. Complex low-level models make it possible to study in detail the characteristics of the readout procedure and the evolution of the quantum state [27-29], however, they may be redundant for statistical reconstruction of the parameters of a quantum system. In our previous work, we performed the construction of a high-level model of measurement of an ion qubit, based on the statistics of photon counts of the PMT [30]. This allowed us to construct fuzzy measurement operators that take into account the false discrimination of the bright and dark states of the ion when using the threshold algorithm. A brief description of this method is given in Section 2.1. In this paper, we propose a new, more advanced method, which is based on accounting for all collected photon statistics and which generalizes a simpler threshold algorithm (Section 2.2). In Section 3, we will show that such a method can significantly improve ion qubits state reconstruction accuracy. In perspective, this approach can be used in any experiment that requires the accumulation of statistics for each individual quantum circuit, including the experimental implementation of NISQ algorithms.

## 2. FUZZY MEASUREMENTS OF A FLUORESCENT QUBIT

The main goal of ion qubit measurements is to extract information about the amplitudes of its state $|\psi\rangle = c_0|0\rangle + c_1|1\rangle$. The measurement is carried out using some auxiliary short-lived level $|m\rangle$ and laser radiation with a frequency $\omega_{0m}$ resonant for a cyclic transition $|0\rangle \rightarrow |m\rangle$ [31]. The measurement of an ion prepared in the ground state $|0\rangle$ will emit a large number of photons, which are then detected by PMT. Thus, a bright fluorescent level $|B\rangle$ can be matched to the state $|0\rangle$. The ion state $|1\rangle$ corresponds to one of the excited metastable long-lived energy levels, functioning as a dark state $|D\rangle$ due to the absence of significant fluorescence during its measurement. In practice, a certain threshold number of samples $k_0$ is introduced. If the number of photons registered during the measurement is greater than or equal to $k_0$, then the state is considered bright. Otherwise, it is considered dark.

The photon statistics corresponding to the registration of the bright state over time $t$ is described by a Poisson distribution with some intensity $\lambda_B$:

$$P_B(k) = \frac{(\lambda_B t)^k}{k!} \exp(-\lambda_B t), \qquad (1)$$

where $k$ is the number of registered photons. It can be shown that the photon statistics corresponding to the registration of the dark state over time $t$ is described by the following finite sum:

$$P_D(k) = \sum_{k_1=0}^{k} \frac{(\lambda_D t)^{k_1}}{k_1!} \exp(-\lambda_D t) P(k-k_1), \qquad (2)$$

where $\lambda_D$ is the total intensity of dark and background counts recorded by the detector due to the imperfection of measuring instruments [32,33], $P(k)$ is the photon number distribution caused the finite lifetime of the metastable level $|1\rangle$:

$$P(k) = \frac{\lambda t \exp(-\lambda t)(\lambda_B t)^k}{((\lambda_B - \lambda)t)^{k+1}} \gamma((\lambda_B - \lambda)t, k+1) + \exp(-\lambda t)\delta_{k0}. \qquad (3)$$

Here $k = 0,1,2,...$; $\lambda = 1/T_1$ - level $|1\rangle$ decay intensity, where $T_1$ is the amplitude relaxation time; $\delta_{k0}$ - Kronecker symbol, $\gamma(x,a)$ - incomplete gamma function:

$$\gamma(x,a) = \frac{1}{\Gamma(a)} \int_0^x z^{a-1} \exp(-z) dz. \qquad (4)$$

Formulas (1) and (2) are obtained using the apparatus of generating functions [34,35]. A more detailed derivation and analysis of these distributions is given in [30].

## 2.1 Threshold measurement model

The presence of the detector dark counts, background noise and amplitude relaxation of the level $|1\rangle$ results in qubit readout errors: with probability $\varepsilon_{10} = \sum_{k=0}^{k_0-1} P_B(k)$ the state $|0\rangle$ will be mistaken for the state $|1\rangle$, and vice versa with probability $\varepsilon_{01} = \sum_{k=k_0}^{\infty} P_D(k)$. Thus, the accuracy and reliability of the qubit measurement is reduced. If the initial probability of detecting a qubit in the state $|0\rangle$ was $p_0$, then the described measurement scheme will give a distorted probability of the form $p_0' = (1-\varepsilon_{10})p_0 + \varepsilon_{01}(1-p_0)$. However, if enough statistical data is accumulated, it is possible to estimate $p_0'$ experimentally and obtain an estimate of the initial probability $p_0$.

It is possible to reconstruct the state with high accuracy even using data which contains significant readout errors. To do that, one needs to construct a fuzzy measurements model [25,26]. In the case under consideration, non-ideal POVM readout operators $\Lambda_0^T = (1-\varepsilon_{10})|0\rangle\langle 0| + \varepsilon_{01}|1\rangle\langle 1|$ and $\Lambda_1^T = \varepsilon_{10}|0\rangle\langle 0| + (1-\varepsilon_{01})|1\rangle\langle 1|$ are introduced, corresponding to the registration of results "0" and "1", respectively. If the basis change unitary transformation $U$ is performed before readout, then effective POVM measurement operators take the following form [30]

$$\Lambda_0^{T,U} = U^\dagger \Lambda_0^T U, \quad \Lambda_1^{T,U} = U^\dagger \Lambda_1^T U. \qquad (5)$$

It is easy to see that this pair of operators also forms a POVM for any unitary $U$. Operators (5) are used in the reconstruction of the quantum state instead of standard projectors. The described method of accounting for readout errors $\varepsilon_{01}$ and $\varepsilon_{10}$ is arising from the non-clear boundary between the bright and dark state. Therefore, we will call it the threshold method.

## 2.2 Photon count measurement model

The threshold method uses stripped-down statistics obtained after discrimination of photon counts. Instead, we propose to use the initial data of the photon counts based on the distributions (1) and (2). Then each number of registered photons will have its own POVM operator:

$$\Lambda_k^{PC} = P_B(k)|0\rangle\langle 0| + P_D(k)|1\rangle\langle 1|, \quad k = 0,1,2,... \qquad (6)$$

Due to the normalized distributions $P_B(k)$ and $P_D(k)$, in total, all these operators will give a single operator, so the given set also forms a POVM. Strictly speaking, it contains an infinite number of elements, but in practice it is sufficient to limit the maximum possible number of recorded samples $n_{ph}$ from above. In addition, we note that the operators $\Lambda_0^T$ and $\Lambda_1^T$ introduced earlier for the threshold method are a reduced version of operators (6). In fact, it is easy to see that $\Lambda_0^T = \sum_{k=k_0}^{\infty} \Lambda_k^{PC}$ and $\Lambda_1^T = \sum_{k=0}^{k_0-1} \Lambda_k^{PC}$. In the case of measuring two qubits, we will assume for simplicity that the measurement of each qubit occurs independently, and the readout errors are identical for them.

Then the fact of registering $k_1$ samples for the first ion and $k_2$ samples for the second corresponds to the measurement operator $\Lambda^{PC}_{k_1,k_2} = \Lambda^{PC}_{k_1} \otimes \Lambda^{PC}_{k_2}$. In the presence of a basis change operation, effective measurement operators are obtained similarly to formula (5): $\Lambda^{PC,U}_{k_1,k_2} = U^\dagger \Lambda^{PC}_{k_1,k_2} U$.

Let $D$ be a set of experimental data obtained as a result of measurements with different $U$. The use of these data in conjunction with the obtained operators makes it possible to estimate the initial quantum state. Below we will consider the maximum likelihood estimation with the root parametrization of the density matrix [26,36]. The software implementation of the corresponding algorithm in MATLAB is available at the link [37]. The direct use of operators $\Lambda^{PC,U}_{k_1,k_2}$ inside the algorithm instead of standard projectors makes it possible to effectively account for readout errors. It is important to note that the model based on counting the number of photons is much more computationally complex than the threshold method. In the case of a model based on counting the number of photons, the total number of measurement operators is $[(n_{ph}+1)n_b]^{n_q}$, where $n_{ph}$ is the maximum number of recorded photons, $n_b$ is the number of measurement bases for each qubit and $n_q$ is the number of qubits in the register. In the case of the threshold algorithm, there are only 2 basis states instead of $(n_{ph}+1)$ and the corresponding total number of operators is $(2n_b)^{n_q}$. For example, let $n_{ph} = 20$, $n_b = 3$, $n_q = 2$. Then the total number of operators is 36 for the threshold algorithm and 3969 for the model based on the count of the number of photons. However, as we will show below, the reward for a significant increase in computational complexity is an increase in the information obtained from measurements. As a result, we observe an increase in the accuracy of the reconstruction of quantum states.

## 3. ANALYSIS OF THE ACCURACY CHARACTERISTICS OF FUZZY QUANTUM MEASUREMENTS

Both described models of quantum measurements allow us to take into account the distributions of the numbers of registered photons. The photon counting model described in this paper, however, uses more information compared to the threshold model and is therefore able to provide higher fidelity of reconstruction of the parameters of the quantum state. To show this, we will perform an analysis of the matrix of complete information describing the statistical fluctuations of the fidelity of reconstruction of the quantum state parameters [36]. In particular, it allows us to calculate the probability distribution $P$ for the value $1 - F$, where $F$ characterizes fidelity between true and reconstructed states. To do this, we use the fact that this random variable has a generalized chi-square distribution in the asymptotic limit on the number of representatives:

$$1 - F \sim \sum_{j=1}^{\nu} d_j \xi_j^2, \tag{7}$$

where $\nu$ is the number of independent parameters that determine the quantum state, $d_j$ are the parameters determined by the matrix of complete information, and $\xi_j \sim N(0,1)$ are independent standard normal random variables. Since the terms in (7) are independent, the characteristic function of the whole sum is determined by the following product:

$$\varphi_{1-F}(t) = \prod_{j=1}^{\nu} \frac{1}{\sqrt{1 - 2id_j t}}. \tag{8}$$

The desired distribution $P(1-F)$ can be obtained numerically by taking the inverse Fourier transform from function (8). A software implementation of this procedure in MATLAB is available at [37]. In calculations, the operators given in Section 2 are substituted as measurement operators.

Consider the case of tomography of a two-qubit state using the Pauli measurement protocol [38]. The parameters of distributions (1) and (2) are: $t = 1$, $\lambda = 0.05$, $\lambda_D = 0.05$, $\lambda_B = 3$. The total sample size is $n = 10^6$. Since the distribution of fidelity loss $P$ depends on the measured state, consider a set of 500 random pure states uniformly distributed by the Haar measure [39,40]. Figure 1 shows the $P$ curves for each of the obtained states. A convenient quantitative assessment of the quality of tomography is the loss function, defined as follows [41]:

$$L = n\langle 1-F \rangle. \tag{9}$$

The average value of the loss function in the example under consideration was $L_{mean} = 5.507 \pm 0.013$, which is noticeably better compared to the threshold method, for which a value of $L_{mean} = 6.020 \pm 0.014$ was obtained under similar conditions.

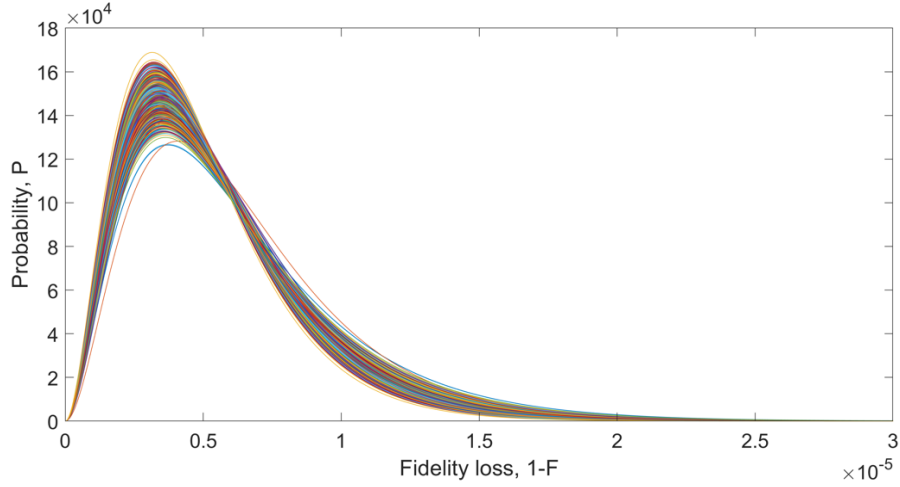

Figure 1. Photon count model. A set of universal fidelity loss distribution curves for 500 pure two-qubit states uniformly distributed by the Haar measure. Sample size is $n = 10^6$. Parameters for $P_D$ and $P_B$ distributions are: $t = 1$, $\lambda = 0.05$, $\lambda_D = 0.05$, $\lambda_B = 3$.

Figure 2a shows a comparison of the photon counting method with the threshold method. The averaged universal fidelity distributions for a sample of 500 pure two-qubit states uniformly distributed by the Haar measure are shown. The solid curve is the photon count method, the dashed curve is the threshold method. It can be seen that the solid curve corresponding to the photon count is narrower (its maximum is correspondingly higher) and shifted to the left compared to the dashed line corresponding to the threshold algorithm. From this it can be concluded that the approach based on counting the number of photons gives more information about the parameters of the state compared to the threshold method.

To verify the obtained distributions, we performed a series of numerical experiments on tomography of random states. In each of them, experimental data were generated from theoretical distributions. These data were further used to reconstruct the quantum state using the root approach and the maximum likelihood method [26,36]. Figure 2b demonstrates a close correspondence between the theory and the numerical experiment on the reconstruction of 500 states uniformly distributed by the Haar measure. Such a close correspondence between theory and experiment indicates the adequacy of the developed method.

Figure 2 corresponds to the sample size $n = 10^6$; parameters for distributions $P_D$ and $P_B$: $t = 1$, $T_1 = 1$; $\lambda_D = 0.01$, $\lambda_B = 6$. The average fidelity losses were: $\langle 1-F \rangle = 1.58 \cdot 10^{-5}$ for photon counting; $\langle 1-F \rangle = 2.61 \cdot 10^{-5}$ for the threshold method.

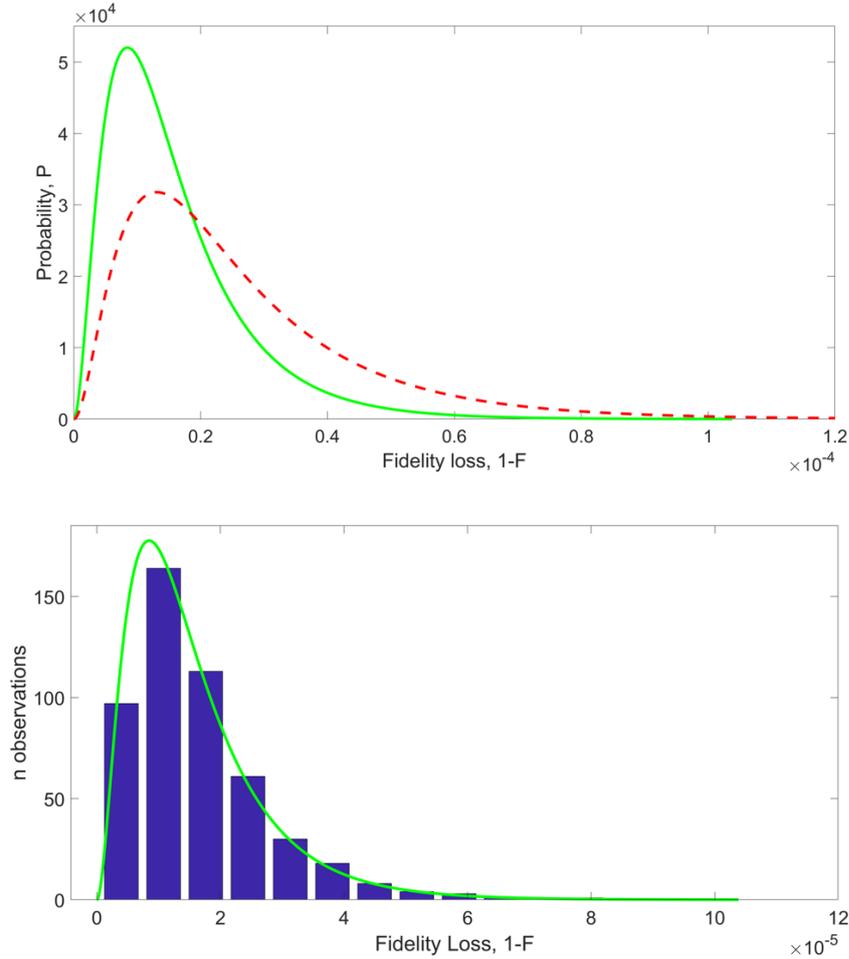

Figure 2. Distribution of fidelity losses. At the top is the comparison of the photon counting method (solid curve) and the threshold method (dashed curve). Below is the comparison of the theory based on the method of counting the number of photons (green curve) with the results of a numerical experiment (histogram) on the reconstruction of 500 random states evenly distributed by the Haar measure.

The duration of measurement in the case of ion qubits directly affects the accuracy of tomography. In the case of small measurement times, the number of fluorescent photons is small, so it is difficult to identify the "0" state (the error $p_{10}$ is large), which leads to the fact that the measurement accuracy is small (accuracy losses are large). However, even in the case of too long measurement times, the accuracy decreases (losses increase), since in this case the role of amplitude (energy) relaxation increases and it becomes difficult to identify the "1" state (the probability of error is high $\varepsilon_{01}$; the ion, which was initially on the upper dark level, manages to fall to the lower bright level and begin to glow, creating the illusion of a bright state). Such reasoning suggests the existence of some optimal measurement time corresponding to minimal loss of fidelity. This is confirmed by the results of numerical calculations, which are presented in Figure 3. The simulation results depend on the products of the intensities and measurement time, i.e. on $\lambda t$, $\lambda_B t$, $\lambda_D t$. In the example under consideration, the parameters $\lambda t$, $\lambda_B t$ and $\lambda_D t$ are fixed, and the time $t$ changes.

In the case of using the threshold method, the minimum of average losses $L_{mean} = 4.919 \pm 0.013$ is achieved at $t = 2.08$, while the average losses for the photon counting method are the minimum of average losses is achieved at $t = 2.23$, while the average losses are $L_{mean} = 4.620 \pm 0.020$. Note that under conditions of perfect distinguishability of bright and dark levels, both methods lead to the same average losses, which are $L_{mean} = 3.26675 \pm 0.00087$. Note also that the theoretically possible minimum fidelity losses for $s=4$-dimensional systems are $L_{\min}=s-1=3$ [41]. To achieve this minimum, however, it is necessary to use projection on the entangled states, which is difficult to implement technologically. Figure 3 provides a clear illustration of the advantages of the method based on counting the number of photons compared to the threshold method. At each point there are 500 random pure two-qubit states, uniformly distributed according to the Haar measure. Spline interpolation was performed between the points.

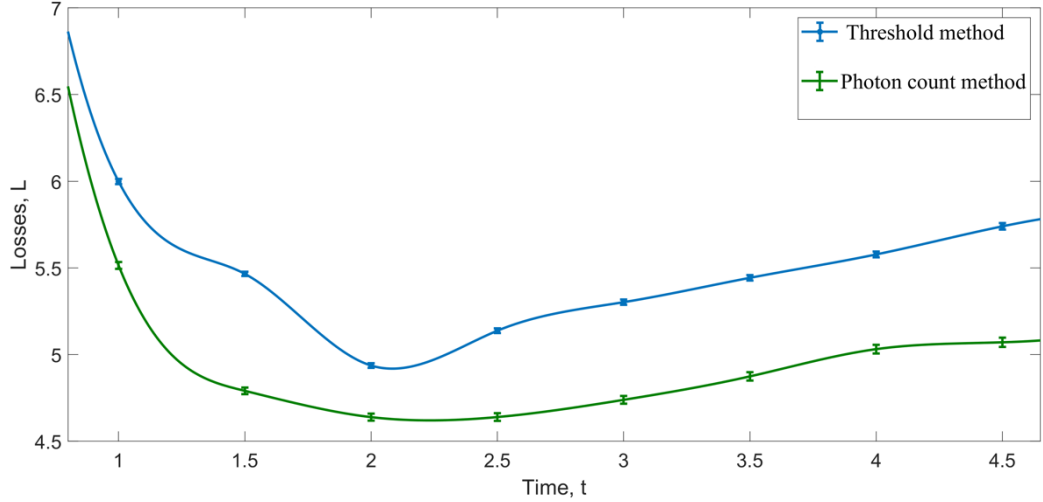

Figure 3. Comparison of the method based on counting the number of photons with the threshold method. The dependence of the fidelity loss on the measurement time *t*, the other parameters are the same as in Figure 1.

Figure 4 illustrates the dependence of fidelity losses on the amplitude relaxation time $T_1$. We see that the advantage of the method based on counting the number of photons over the threshold method is especially pronounced in conditions of strong amplitude relaxation, when the parameter $T_1$ is small. So, for the $T_1=1$ method of counting the number of photons, the average losses are $L_{mean}=15.97\pm0.18$, and for the threshold method $L_{mean}=25.81\pm0.19$. In this case, the threshold method is about 1.6 times less accurate than the method of counting the number of photons. On the other hand, with $T_1=100$ the losses are $L_{mean}=3.5451\pm0.0091$ for the photon count method and $L_{mean}=3.707\pm0.010$ for the threshold method. In this case, the threshold method is only 1.046 times less accurate than the photon count method.

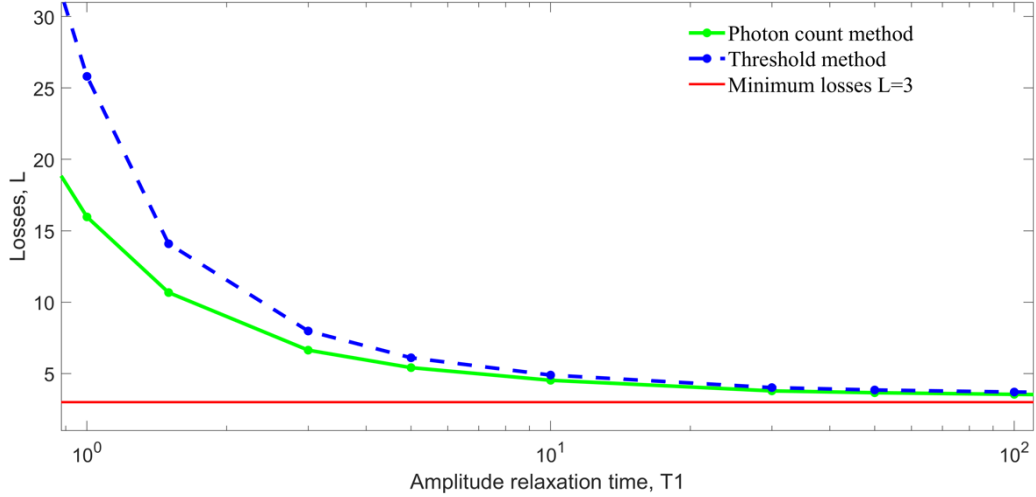

Figure 4. Dependence of fidelity losses on the amplitude relaxation time. The horizontal scale is represented on a logarithmic scale. The solid green line corresponds to the method based on counting the number of photons, the dashed line corresponds to the threshold method, the horizontal red line corresponds to the theoretically possible minimum losses $L_{\min} = 3$. At each point there are 200 random pure two-qubit states, uniformly distributed by the Haar measure. Parameters for distributions $P_D$ and $P_B$ are: $t = 1$, $\lambda_D = 0.01$, $\lambda_B = 6$.

## 4. CONCLUSIONS

We have proposed a method of quantum tomography of ion qubits based on the analysis of the total number of photon counts obtained as a result of observing the fluorescence of the bright state. We compared this method with the previously proposed threshold method, which uses only the information about the number of counts greater than or equal to a certain threshold, but at the same time takes into account the probability of a false determination of the state of the qubit. It has been shown that the photon counting method, although computationally more complex, carries more information about the quantum state and allows for higher tomography accuracy in comparison with the threshold method.

In the absence of other systematic errors, the fuzzy measurement approach makes it possible to ensure the quantum state reconstruction accuracy close to 100% using a sufficiently large set of statistical data. At the same time, the model of standard projection measurements does not take into account the influence of readout errors, which leads to systematic errors that significantly limit the maximum possible accuracy of state reconstruction.

## 6. ACKNOWLEDGEMENTS

The work was carried out within the project No. 22-12-00263 of the Russian Science Foundation.

Section 2 "Fuzzy measurements of a fluorescent qubit" was supported by Theoretical Physics and Mathematics Advancement Foundation "BASIS" (Grant No. 20-1-1-34-1).